\documentclass[12pt]{article}

\usepackage{amsmath}
\usepackage{amssymb}
\usepackage{graphicx}
\usepackage{float}
\usepackage[dvips]{color}
\usepackage{colordvi}
\usepackage{url}

\begin{document}

\baselineskip=18pt

\def\gap#1{\vspace{#1 ex}}
\def\be{\begin{equation}}
\def\ee{\end{equation}}
\def\bal{\begin{array}{l}}
\def\ba#1{\begin{array}{#1}}  
\def\ea{\end{array}}
\def\bea{\begin{eqnarray}}
\def\eea{\end{eqnarray}}
\def\beas{\begin{eqnarray*}}
\def\eeas{\end{eqnarray*}}
\def\del{\partial}
\def\eq#1{(\ref{#1})}
\def\fig#1{Fig \ref{#1}} 
\def\re#1{{\bf #1}}
\def\bull{$\bullet$}
\def\nn{\\\nonumber}
\def\ub{\underbar}
\def\nl{\hfill\break}
\def\ni{\noindent}
\def\bibi{\bibitem}
\def\ket{\rangle}
\def\bra{\langle}
\def\vev#1{\langle #1 \rangle} 
\def\lsim{\stackrel{<}{\sim}}
\def\gsim{\stackrel{>}{\sim}}
\def\mattwo#1#2#3#4{\left(
\begin{array}{cc}#1&#2\\#3&#4\end{array}\right)} 
\def\tgen#1{T^{#1}}
\def\half{\frac12}
\def\floor#1{{\lfloor #1 \rfloor}}
\def\ceil#1{{\lceil #1 \rceil}}

\def\mysec#1{\gap1\ni{\bf #1}\gap1}
\def\mycap#1{\begin{quote}{\footnotesize #1}\end{quote}}

\def\bit{\begin{item}}
\def\eit{\end{item}}
\def\benu{\begin{enumerate}}
\def\eenu{\end{enumerate}}
\def\a{\alpha}
\def\as{\asymp}
\def\ap{\approx}
\def\b{\beta}
\def\bp{\bar{\partial}}
\def\cA{{\cal{A}}}
\def\cD{{\cal{D}}}
\def\cL{{\cal{L}}}
\def\cP{{\cal{P}}}
\def\cR{{\cal{R}}}
\def\da{\dagger}
\def\de{\delta}
\def\tD{\tilde D}
\def\e{\eta}
\def\ep{\epsilon}
\def\eqv{\equiv}
\def\f{\frac}
\def\g{\gamma}
\def\G{\Gamma}
\def\h{\hat}
\def\hs{\hspace}
\def\i{\iota}
\def\k{\kappa}
\def\lf{\left}
\def\l{\lambda}
\def\la{\leftarrow}
\def\La{\Leftarrow}
\def\Lla{\Longleftarrow}
\def\Lra{\Longrightarrow}
\def\L{\Lambda}
\def\m{\mu}
\def\na{\nabla}
\def\nn{\nonumber\\}
\def\mm{&&\kern-18pt}  
\def\om{\omega}
\def\O{\Omega}
\def\P{\Phi}
\def\pa{\partial}
\def\pr{\prime}
\def\r{\rho}
\def\ra{\rightarrow}
\def\Ra{\Rightarrow}
\def\ri{\right}
\def\s{\sigma}
\def\sq{\sqrt}
\def\S{\Sigma}
\def\si{\simeq}
\def\st{\star}
\def\t{\theta}
\def\ta{\tau}
\def\ti{\tilde}
\def\tm{\times}
\def\tr{\textrm}
\def\Tr{{\rm Tr}}
\def\T{\Theta}
\def\up{\upsilon}
\def\Up{\Upsilon}
\def\v{\varepsilon}
\def\vh{\varpi}
\def\vk{\vec{k}}
\def\vp{\varphi}
\def\vr{\varrho}
\def\vs{\varsigma}
\def\vt{\vartheta}
\def\w{\wedge}
\def\z{\zeta}

\thispagestyle{empty}
\addtocounter{page}{-1}
{}
\vskip-5cm
\begin{flushright}
TIFR/TH/11-24\\
CCTP-2011-14\\
\end{flushright}
\vspace*{0.1cm} \centerline{\Large \bf Gregory-Laflamme as the
confinement/deconfinement}
\vspace*{0.1cm}
 \centerline{\Large \bf  transition in holographic QCD}
\vspace*{1 cm} 
\centerline{\bf 
Gautam~Mandal$^1$ and Takeshi~Morita$^2$}
\vspace*{0.5cm}
\centerline{\rm $^1$\it Department of Theoretical Physics,}
\centerline{\it Tata Institute of Fundamental Research,} 
\centerline{\it Mumbai 400 005, \rm INDIA}
\vspace*{0.3cm}
\centerline{\rm $^2$\it Crete Center for Theoretical Physics}
\centerline{\rm \it Department of Physics}
\centerline{\it University of Crete, 71003 Heraklion, \rm Greece}
\vspace*{0.3cm}
\centerline{\tt email: mandal@theory.tifr.res.in, takeshi@physics.uoc.gr
}

\vspace*{0.4cm}
\centerline{\bf Abstract}
\vspace*{0.3cm} 

\enlargethispage{1000pt}

 We discuss the phase structure of $N$ D4 branes wrapped on a
  temporal (Euclidean) and a spatial circle, in terms of  the
  near-horizon geometries.  This system has been studied previously
  to understand four dimensional pure $SU(N)$ Yang-Mills theory (YM4)
  through holography. In the usual treatment of the subject, the phase
  transition between the solitonic D4 brane  and the black D4 brane is
  interpreted as the strong coupling continuation of the
  confinement/deconfinement transition in YM4.  
    We show that this interpretation is not valid, since the black D4 brane and the deconfinement phase of YM4 have different realizations of the $Z_N$ centre symmetry and cannot be identified.
 We propose an alternative gravity dual of the
  confinement/deconfinement transition in terms of a Gregory-Laflamme
  transition of  the soliton  in the IIB frame, where the strong
  coupling continuation of the deconfinement phase of YM4 is a
  localized D3 soliton.  Our proposal offers a new explanation of
  several aspects of the thermodynamics of holographic QCD.  As an
  example, we show a new mechanism of chiral symmetry restoration
  in the Sakai-Sugimoto model.  The issues discussed in this paper
  pertain to gravity duals of non-supersymmetric gauge theories in
  general. 
\newpage


\tableofcontents

\section{Introduction}

Strongly coupled field theories such as QCD and various condensed
matter systems are not amenable to perturbative calculations and
require numerical or other non-perturbative tools. A powerful new tool
in this context is holography, which maps gauge theories to gravity in
a higher dimension \cite{Maldacena:1997re, Itzhaki:1998dd, Gubser:1998bc, Witten:1998qj, Witten:1998zw, Aharony:1999ti}.
 Holography, however, is most reliable and quantitative for special
supersymmetric gauge theories. In the generic case, including for pure
Yang Mills theories, inferences from gravity  tend to be
less quantitative due to various degrees of extrapolations, although
such exercises have proved to be rather valuable, e.g., in the context
of holographic QCD \cite{Witten:1998zw, Aharony:1999ti, Gross:1998gk, 
  Kruczenski:2003uq, Sakai:2004cn}.

Recently, it was pointed out in \cite{Mandal:2011hb} that there are
problems with a naive application of holography to a two dimensional
large $N$ bosonic gauge theory at finite temperature.  The dual of
this system is based on gravity backgrounds involving $N$ D2 branes on
$T^2$.  The phase diagram of gravity, interpreted naively, does not
admit a unique continuation to the regime of gauge theory, owing to its
dependence on the boundary condition for fermions on the brane. However, it
was shown in \cite{Mandal:2011hb} that the phase structures in gravity
and gauge theory 
could be smoothly connected  through an appropriate choice
of fermion boundary conditions.

In this paper, we will show that similar problems arise when we
holographically analyse more general non-supersymmetric gauge theories
at finite temperature. The particular example we will focus on is that
of holographic QCD from D4 branes \cite{Witten:1998zw, Aharony:1999ti, Gross:1998gk, 
  Kruczenski:2003uq, Sakai:2004cn}\footnote{We will briefly discuss the generalization to a $p$ dimensional bosonic gauge theory and its holographic dual in section \ref{Conclusions}.}. We will discuss some
problems  with
the usual correspondence between the
confinement/deconfinement transition in QCD and the Scherk-Schwarz
transition between  a solitonic D4 brane  and a black D4 brane in the gravity dual.
Some of these problems were first discussed in \cite{Aharony:2006da}
(see section \ref{sec-AP} for further details).
Especially we will show that the black D4 brane cannot be identified
with the (strong coupling continuation of the) deconfinement phase in QCD
in four dimensions. 
As a resolution of these problems, we will
propose an alternative scenario in which the confinement/deconfinement
transition corresponds to a Gregory-Laflamme transition
\cite{Gregory:1994bj, Aharony:2004ig, Harmark:2004ws} between a uniformly distributed  D3 soliton  and
a localized D3 soliton in the IIB frame.

The scenario we propose suggests that we need to reconsider several
previous results in holographic QCD including the Sakai-Sugimoto model
\cite{Sakai:2004cn}.  One important ingredient in the Sakai-Sugimoto
model is the mechanism of chiral symmetry restoration at high
temperatures \cite{Aharony:2006da}.  We will propose a new mechanism
for chiral symmetry restoration in our framework.

The plan of the paper is as follows. 

Section \ref{sec-review} contains a short review of finite temperature
QCD and the holographic approach to four dimensional Yang Mills theory
(YM4)\footnote{\label{ftnt:ym4} Throughout this paper
we will often use the notation YM4 for four dimensional Yang Mills
theory and SYM5 for five dimensional super Yang Mills theory.} using D4 branes. In Section \ref{sec:T} we discuss the gravity theory
at a finite temperature and a `Scherk-Schwarz' transition between a
solitonic D4 brane and a black D4 brane. In Section \ref{sec-AP} we
recall the conventional correspondence between this transition and the
confinement/deconfinement transition of the Yang Mills theory, and
discuss problems with this correspondence, with special emphasis on
the mismatch of $Z_N \times Z_N$ centre symmetry between the black D4
phase and the deconfined 4 dimensional YM theory.  In Section
\ref{sec-GL}, we discuss the phase structure of the D4 branes using an
unconventional (periodic) fermion boundary condition along the
Euclidean time circle and show that the high temperature in gravity is
described by a localized solitonic D3 brane  (or its T-dual) whose
centre symmetry precisely matches with the deconfinement phase of four
dimensional YM theory. This leads us, in Section \ref{sec-GL-CD}, to
propose a new representation of the confinement/deconfinement phase
transition in terms of a GL transition between the solitonic D4 and
the (T-dual of) the localized solitonic D3 (see Table
\ref{table-proposal}). In Section \ref{sec-new}, we show some new
correspondences between phenomena in QCD and their counterparts
in gravity following our proposal. In Section \ref{sec-chiral}, we
suggest a new mechanism of chiral symmetry restoration in the
Sakai-Sugimoto model in keeping with our proposal. Section
\ref{Conclusions} contains the concluding remarks and some open
problems.  

\section{Review of QCD and its gravity dual}
\label{sec-review}

\subsection{Finite temperature QCD}
\label{sec-QCD}

In this section, we briefly review some salient properties of four
dimensional $SU(N)$ pure Yang-Mills theory at a finite temperature at
large $N$, a system which we will subsequently investigate through
holography.

An important symmetry in this theory is the $Z_N$ symmetry along the temporal cycle, which is the centre of the $SU(N)$ symmetry.
The order parameter of this symmetry is the temporal Polyakov loop operator
\begin{align} 
 W_0= \frac1N {\rm Tr } P e^{i \int_0^{\beta} A_0 dx^0 }.
\label{Polyakov}
\end{align} 
If this operator is zero, the $Z_N$ symmetry is preserved and, if not, it is broken.
Especially, if the $Z_N$ symmetry is preserved, physical quantities do not depend on the temporal radius $\beta/2\pi$ at $O(N^2)$ order due to `large $N$ volume
independence' \cite{Eguchi:1982nm, Gocksch:1982en, Kovtun:2007py}.
If the gauge theory is on a torus $S^1_{L_1} \times S^1_{L_2}\times\cdots $, the large $N$ volume independence is generalized and physical quantities do not depend on $L_i$, if the $Z_N$ symmetry along the $i$-th direction is preserved \cite{Narayanan:2007fb}. 

Let us consider the phase structure of the Yang-Mills theory. 
At low temperatures, the confinement phase is the dominant
thermodynamic phase.  In this phase, the $Z_N$ symmetry is
preserved ($W_0=0$). 
 At high temperatures, the deconfinement phase, in which the $Z_N$
symmetry is broken ($W_0\neq 0$), dominates. There is a confinement/deconfinement
transition at an intermediate temperature.  From studies in large
$N$ lattice gauge theories, this is expected to be a first order phase
transition \cite{Lucini:2003zr, Kiskis:2005hf, Panero:2009tv, Datta:2010sq}.

As we remarked above, the free energy in the confinement phase does
not depend on the temperature through the `large $N$ volume independence'.
It implies that the entropy becomes zero at $O(N^2)$. On the other hand, the entropy in the deconfinement phase would be $O(N^2)$.
This has a simple physical interpretation: in the confinement phase, the
spectrum excludes gluon states, which have $O(N^2)$ degree of the
freedom, and consists only of gauge singlet states like glueballs,
leading to an $O(1)$ entropy.

\subsection{Holographic QCD}
\label{sec-holographicQCD}

In this section, we will review the construction of four dimensional
$SU(N)$ pure Yang-Mills theory from $N$ D4 branes
\cite{Witten:1998zw}.  Let us first consider a 10 dimensional
Euclidean spacetime with an $S^1$ and consider D4 branes wrapping on
the $S^1$. 
We define the coordinate along this $S^1$ as $x_4$ and its periodicity as $L_4$.
The effective theory on this brane is a 5 dimensional supersymmetric Yang-Mills theory on the $S^1_{L_4}$.
For the fermions on the brane, the boundary condition along the circle can
be AP (antiperiodic) or P (periodic); to specify the theory, we must
pick one of these two boundary conditions\footnote{\label{AP} Defining a spinor field on a manifold
  $M$ requires specifying a ``spin structure'', which means specifying
  a consistent choice of the sign of the spinorial transformation
  (recall that, given a Lorentz transformation $\L$ on vectors, the
  corresponding spinor transformation $S(\L)$ is ambiguous up to a
  sign $S(\L) \leftrightarrow - S(\L)$). On a simply connected
  Euclidean manifold $M$ there is a unique spin structure; all cycles
  are contractible and the fermion b.c. along any such cycle is
  antiperiodic. If $M$ has the topology $S^1 \times N$ where $N$ is
  simply connected (generalizations to multiple $S^1$ factors are obvious), then the spin structure on $M$ has a two-fold
  ambiguity, where a spinor can be either periodic (P) or
  anti-periodic (AP) along the $S^1$. See, e.g. \cite{AMP}. For
  examples relevant to this paper, see \cite{Witten:1998zw}).}. Let
us take the AP boundary condition.  This gives rise to fermion masses
proportional to the Kaluza-Klein scale $1/L_4$, leading to
supersymmetry breaking (this is called the SS --- Scherk-Schwarz---
mechanism).  This, in turn, induces masses for the adjoint scalars and for
$A_4$, which are proportional to $\lambda_4/L_4$ at
one-loop. Therefore, if $\lambda_4$ is sufficiently small and the
dynamical scale $\L_{YM}$ is much less than both the above mass
scales\footnote{\label{small-lambda} Recall that $\L_{YM} = \L
  \exp[-1/(b\l_4)],$ where $\l_4$ is the value of the coupling at some
  cut-off scale $\L$ and $b$ is the first coefficient of the $\beta$-function; if $\l_4$ is sufficiently small it is possible
  to ensure $\L_{YM} \ll \l_4/L_4,1/L_4$ since as $\l_4$ goes to zero,
  $\exp[-1/(b\l_4)]$ goes to zero faster than $\l_4$.}, then the fermions
and adjoint scalars are decoupled and the 5 dimensional supersymmetric Yang-Mills theory is reduced to a four dimensional pure Yang-Mills theory\footnote{In this construction of the 4 dimensional Yang-Mills theory from the 5 dimensional SYM through the Kaluza-Klein reduction, we have implicitly assumed that the long string modes are suppressed. We will come back to this problem later on.}.

By taking the large $N$ limit of this system {\em a la} Maldacena at low temperatures, we obtain the dual gravity description of the compactified 5 dimensional SYM theory \cite{Witten:1998zw}, which consists, at low temperatures, of a solitonic D$4$ brane solution wrapping the
$S^1_{L_4}$.
  The explicit metric is given by 
\begin{align}
ds^2 =& \a' \left[\frac{u^{3/2}}{\sqrt{d_4 \lambda_5 }}
\left( dt^2  + \sum_{i=1}^{3} 
dx_i^2+f_4(u) dx_4^2 \right)+ \frac{\sqrt{d_4 \lambda_5 }}{u^{3/2}}\left(  
\frac{du^2}{f_4(u)} 
+ u^2 d\Omega_{4}^2 \right)  \right], \nn
& f_4(u)=1-\left( \frac{u_0}{u}\right)^3 , \quad e^{\phi}=\frac{\lambda_5}{(2\pi)^2N}
\left(\frac{u^{3/2}}{\sqrt{d_5 \lambda_5}}  \right)^{1/2}  .
 \label{metric-SD4}
\end{align}
There is also a non-trivial value of the five form
potential which we do not write explicitly. 
  Here $\lambda_5=\lambda_4
L_4$ is the YM coupling on the D$4$ world-volume and $d_4$ is found
by putting $p=4$ in the general formula
\begin{align} 
d_p&= 2^{7-2p}\pi^{(9-3p)/2}\Gamma\left( \frac{7-p}2 \right). 
\label{dp}
\end{align} 

Since the $x_4$-cycle shrinks to zero at $u=u_0$, in order to avoid
possible conical singularities we must choose the asymptotic radius
$L_4$ as follows 
\begin{align}
\frac{L_4}{2\pi}= \frac{\sqrt{d_4 \l_{5}}}{3} u_0^{-1/2} .
\label{u0-beta}
\end{align} 
With this choice, the contractible $x_4$-cycle, together with the
radial direction $u$, forms a so-called `cigar' geometry which is
topologically a disc. In order that the fermions are well-defined on
this geometry, they must obey the AP boundary
condition. The AP boundary condition along $x_4$ is, of course, consistent with the
choice of fermion b.c. in the boundary theory.

The leading order gravity solution described above is not
always valid. E.g. in order that the stringy modes can be ignored, we
should ensure that the curvature in string units must be small.  In
other words, the typical length scale of this solution near $u=u_0$,
{\em viz}, $l= \left(\alpha' \sqrt{d_4 \lambda_{5}}
u_0^{1/2}\right)^{1/2}$, must satisfy $l\gg \sqrt{\alpha'}$
\cite{Itzhaki:1998dd}. This condition turns out to be equivalent to
\begin{align} 
\lambda_4 \gg 1 .
\label{gravity-cond-SD4}
\end{align} 
This is opposite to the condition which we found for the validity of
the four-dimensional gauge theory description (see footnote \ref{small-lambda}).
Thus, this gravity solution can describe the 5 dimensional SYM but cannot directly describe the 4 dimensional YM theory, inferences about which can only be obtained through extrapolation.
This is a common problem in the construction of holographic duals of non-supersymmetric gauge theories. This has been discussed at length; in particular, the
gravity description, which necessitates extrapolation to strong
coupling, has been likened  (cf. \cite{Gross:1998gk} and \cite{Aharony:1999ti}, p. 196-197) to
strong coupling lattice gauge theory.
Many interesting results, including the qualitative predictions in \cite{Witten:1998zw, Aharony:1999ti, Gross:1998gk, Sakai:2004cn}, have been obtained using this prescription.

Let us explore some properties of the soliton solution (\ref{metric-SD4}).
This solution is expected, on the basis of several arguments, to correspond to the confinement phase in
the four dimensional gauge theory.  For
example, it can be shown that the classical action of this solution at
a finite temperature $1/\b$ is \cite{Harmark:2004ws, Horowitz:1998ha, Harmark:2007md}\footnote{In Section \ref{sec:T} we will describe more details of holographic QCD at finite temperatures. Eq. \eq{action-SD4}
describes the stable phase of the system only at a sufficiently low temperature.} 
\begin{align} 
S_{SD4}/N^2V_{3}=-C_4 \lambda_4 \beta L_4^{-4},
\label{action-SD4}
\end{align} 
where $V_3$ is the volume of the three spatial dimensions ($x_1,x_2,x_3$) 
and $C_4$ is a constant which is obtainable from the formula below \cite{Itzhaki:1998dd, Mandal:2011hb} by putting $p=4$ 
\begin{align} 
C_p&=\frac{5-p}{2^{11-2p}\pi^{(13-3p)/2}\Gamma((9-p)/2) a_p^{2(7-p)/(5-p)}} , 
\quad
a_p=\frac{7-p}{4\pi d_p^{1/2}}.
\end{align} 
Thus the free energy of this solution, $F= S_{SD4}/\beta$, is
independent of temperature, and hence the entropy is zero at $O(N^2)$
order. These facts are consistent with the interpretation of this
solution as the confinement phase in the large $N$ gauge theory, in which
 the temporal $Z_N$ symmetry is preserved.

Note that the action (\ref{action-SD4}) divided by $L_4$ does depend on $L_4$, implying that the $Z_N$ symmetry along the $x_4$-cycle is broken. Although  this observation might appear irrelevant for the 4 dimensional Yang-Mills theory, it is this symmetry breaking which allows a conventional KK reduction to four dimensions (see 
the next two sections for more details).

\section{ Finite temperature holographic QCD }
\label{sec-problem}

\subsection{\label{sec:T} Confinement/deconfinement transition in 
Holographic QCD in the standard scenario}

In existing holographic studies of the thermodynamics in the Yang-Mills theory, it is assumed that the black D4 brane solution corresponds to the deconfinement phase and the Scherk-Schwarz transition in the gravity is related to the confinement/deconfinement transition.  In this
section, we review this scenario and we will explain its problems in
section \ref{sec-AP}.

To discuss holographic QCD at finite
temperatures, we begin by compactifying Euclidean
time in the boundary theory on a circle with periodicity $\beta=1/T$. To
permit description in terms of four dimensional gauge theory, the temperature
should, of course, be kept far below the KK scale, $T \ll \lambda_4/L_4$ \footnote{The precise conditions for the decoupling of KK modes from YM4 are $T \ll 1/L_4$ for the fermions and $T \ll \lambda_4/L_4$ for the bosonic modes (with $\lambda_4 \ll 1$). Since the second condition includes the first one at weak coupling,  we omit the first condition.}.
In order to determine the gravitational theory, we need to fix the
periodicity of fermions in the gauge theory along the time cycle.  In
the existing literature, the fermion boundary condition is chosen to
be AP according to the usual practice for thermal fermions (we will
see in section \ref{sec-phase} that this choice is not imperative
here). The resulting fermionic bc's\footnote{\label{ftnt-top} Here and elsewhere the boundary
  conditions will always refer to those of the boundary theory along
  $S^1_\beta \times S^1_{L_4}$, respectively.}: (AP, AP) along $(t, x_4)$ lead to
a $Z_2$ symmetry of the system under $t \leftrightarrow x_4$. In the
gravity dual, we must therefore, include the $Z_2$-transform of the solution
\eq{metric-SD4}, which is a black D4 solution:
\begin{align}
ds^2 =& \a' \left[\frac{u^{3/2}}{\sqrt{d_4 \lambda_5 }}
\left( f_4(u)dt^2  + \sum_{i=1}^{3} 
dx_i^2+ dx_4^2 \right)+ \frac{\sqrt{d_4 \lambda_5 }}{u^{3/2}}\left(\frac{du^2}{f_4(u)} 
+ u^2 d\Omega_{4}^2 \right)  \right], \nn
& f_4(u)=1-\left( \frac{u_0}{u}\right)^3. 
\label{metric-BD4}
\end{align}
Here $u_0$ in this metric is related to $\beta$ as in (\ref{u0-beta}) 
\begin{align}
\frac{\beta}{2\pi}= \frac{\sqrt{d_4 \l_{5}}}{3} u_0^{-1/2}.
\label{smooth-cond-BD4}
\end{align} 
This solution has a contractible $t$-cycle and the fermions must
obey the anti-periodic boundary condition (see footnote \ref{AP}),
consistent with the AP b.c. in the gauge theory.
Analogously to (\ref{gravity-cond-SD4}), the
gravity description of this solution is reliable if $\lambda_4 \gg
\beta/L_4$\footnote{In several equations for the
  black D4 brane in this paper, we use the four dimensional gauge
  coupling $\lambda_4=\lambda_5/L_4 $, which obscures the $\beta
  \leftrightarrow L_4$ symmetry. If we write these equations using
  $\lambda_5$ instead of $\lambda_4$, we will see that they are rather
  simply obtained from the corresponding equations for the solitonic
  D4 brane via the exchange $\beta \leftrightarrow L_4$.}. 

 The classical action density can be evaluated, as before,
yielding
\begin{align} 
S_{BD4}/N^2V_{3}=-C_4 \lambda_4 L_4^2 \beta^{-5}.
\label{action-BD4}
\end{align} 
In contrast with \eq{action-SD4}, the
free energy now is a function of $\b$ and hence the entropy  is $O(N^2)$ which is appropriate for a description of the
deconfinement phase in the gauge theory.

By comparing (\ref{action-SD4}) and (\ref{action-BD4}), we see that at
low temperatures (large $\b$) the solitonic D4 solution dominates,
while at high temperatures the black brane dominates. The transition
between the two solutions, which we will call the ``Scherk-Schwarz
(SS) transition'', occurs at \cite{Horowitz:1998ha}
\begin{align}
\beta_{SS}=  L_4 .
\label{SS-4}    
\end{align}
In the standard discussions of holographic QCD 
  \cite{Aharony:1999ti, Kruczenski:2003uq, Aharony:2006da}, this transition is
interpreted as the (continuation of the) confinement/deconfinement transition in the gauge
theory.  In the following sections, we will explain various problems
with this interpretation and present a resolution by proposing an
alternative scenario.

\subsection{Problems with the standard gravity analysis}
\label{sec-AP}

The previous subsection describes the usual relation between the confinement/deconfinement transition in 4 dimensional Yang-Mills theory and the SS transition in holographic QCD.
However as we will now discuss, there are several problems\footnote{The problems discussed in this
section are not about the use of holography in predicting the phase diagram  of strongly coupled 5 dimensional SYM, but rather about its extrapolation to four dimensional Yang-Mills theory.} associated with this correspondence. Some of these problems were mentioned earlier
in \cite{Aharony:2006da}.

\begin{figure}
\begin{center}
\includegraphics[scale=.75]{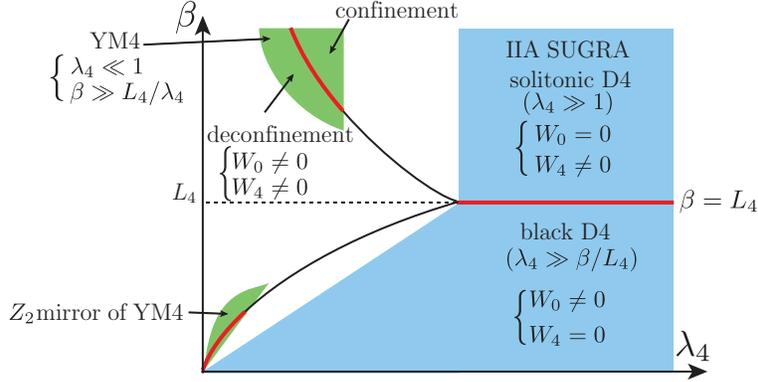}
\caption{The phase structure of the five dimensional SYM  (SYM5) on $S^1_\beta \times S^1_{L_4} $
  with the (AP,AP) boundary condition.
  The gravity analysis is valid in the strong coupling region (the blue region).
The 4 dimensional YM description is valid in the upper green regime.
The lower green regime is the mirror of the upper one via the $Z_2$ symmetry $\beta \leftrightarrow L_4$. The solid red line in the green regions
depicts the known confinement/deconfinement transitions in gauge theory and its $Z_2$ mirror. The solid red line in the blue region is the SS transition
\eq{SS-4}. These lines demarcate phases characterized by the $Z_N \times Z_N$
order parameters $W_0$ and $W_4$ (see Table \ref{table-solutions}). The solid black lines correspond to a minimal extrapolation of the phase boundaries through the intermediate region. The dotted line denotes another possible phase transition which is allowed by the $Z_2$ symmetry but across which the $Z_N \times Z_N$ symmetry does not change.}
\label{fig-D4-AP-phase}
\end{center}
\end{figure}

In order to understand these problems, it is convenient to consider
first the phase structure of the 5 dimensional SYM on the $S^1_\beta
\times S^1_{L_4} $. See Figure \ref{fig-D4-AP-phase}.  As we have
seen, the gravity description is valid in the strong coupling regime
of this theory (the blue region in Figure \ref{fig-D4-AP-phase}) which
is characterized by the solitonic D4 solution and the black D4
solution.  On the other hand, the 4 dimensional Yang-Mills theory is
realized at weak coupling ($\lambda_4 \ll 1$) and low temperature
($\beta \gg L_4/\lambda_4$) (the upper green region in Figure
\ref{fig-D4-AP-phase}) which is characterized by the confinement phase
and the deconfinement phase.  The usual proposal is that the solitonic D4 and
black D4 solutions correspond to the confinement and deconfinement
phases respectively.

To examine this proposal in detail, it is useful to consider the
realization of the
$Z_N \times Z_N$ centre symmetry in various 
phases of the five-dimensional SYM theory
along $t$ and $x_4$. The $Z_N$ symmetry along $t$ is characterized
by the order parameter $W_0$ which is the temporal Polyakov loop
\eq{Polyakov}. Similarly the $Z_N$ symmetry along $x_4$ is
characterized by the order parameter $W_4$ which is the Polyakov loop
around the $x_4$-cycle:
\begin{align} 
W_4= \frac1N {\rm Tr } P e^{i \int_0^{L_4} A_4 dx^4 }.
\label{w4}
\end{align}  
In Table \ref{table-solutions} we have collected values of $(W_0,
W_4)$ for the phases of 5D SYM discussed in this paper\footnote{One cannot, {\em a priori}, rule out additional phases in
  the intermediate coupling regime (the white region in Figure
  \ref{fig-D4-AP-phase} and \ref{fig-D4-phase-P}.). However, the
  lattice super Yang Mills study in \cite{Catterall:2010fx} has
  analyzed such intermediate coupling regimes in the context of D1
  branes compactified on a periodic circle, and has not found any
  novel phases. This result may indicate that there are no other
  phases in the present case with D4 branes as well.}.

\begin{table}
\begin{center}
\begin{tabular}{l|c|c}
`Confinement' phase in YM4 & $W_0=0$, $W_4 \neq 0$ & AP/P \\
`Deconfinement' phase in YM4 & $W_0 \neq 0$, $W_4 \neq 0$ & AP/P \\
$Z_2$ mirror of `confinement' phase in YM4  & $W_0 \neq 0$, $W_4 = 0$ & AP \\
$Z_2$ mirror of `deconfinement' phase in YM4  & $W_0 \neq 0$, $W_4 \neq
 0$ & AP \\
Solitonic D4 phase & $W_0=0$, $W_4 \neq 0$ & AP/P  \\ 
Black D4 phase & $W_0 \neq 0$, $W_4 = 0$ & AP \\
Localized solitonic D3 phase & $W_0 \neq 0$, $W_4 \neq 0$ & P \\ 
\end{tabular}
\caption{The values of the Polyakov loops in some phases of the 5
  dimensional SYM (SYM5) on  $S^1_\beta \times S^1_{L_4} $ (focussing
on features related to the physics of the four dimensional YM theory).
  The boundary condition for the fermion along $S^1_\beta $ is shown in the right column. (We have fixed the AP b.c. along the $S^1_{L_4}.$)
   The `confinement' phase in YM4 here refers to the phase of SYM5 whose KK
  compactification, at small enough $L_4$, coincides with the usual
  confinement phase of YM4. Similarly, the `deconfinement' phase in YM4 here
  refers to the phase of SYM5 whose KK compactification, at small
  enough $L_4$, coincides with the usual deconfinement phase of
  YM4. The phases labelled by gravity solutions, valid at suitable
  parameter regimes, denote the respective phases of SYM5 at those
  parameter regimes.}
\label{table-solutions} 
\end{center}
\end{table}

Before testing the above proposal for holographic QCD in terms of
these order parameters, let us see how to obtain this table.  The
`confinement' phase denotes a phase of the 5D SYM whose KK
compactification, at small enough $L_4$, coincides with the usual
confinement phase of YM4. The latter, clearly has $W_0 = 0$.  Now,
the fact that we have obtained the 4D phase through a KK reduction,
along $L_4$, implies that the corresponding phase of the 5D theory
must have $W_4 \ne 0$. This follows from the fact
\cite{Kovtun:2007py, Aharony:2005ew} that in a $W_4 =0$ phase, KK
reduction does not work since the effective KK scale in a large $N$
gauge theory becomes $1/(NL_4)$ in stead of the usual $1/L_4$, and
goes to zero at large $N$ \footnote{The fractional KK scale, related
  to the `long string' modes of \cite{Dijkgraaf:1997vv}, can be
  understood as arising from mode shifts of charged fields in the
  presence of Wilson lines whose eigenvalues are uniformly distributed
  along a circle.}. Thus, the 'confinement' phase must have $W_0=0,
W_4 \ne 0$, as shown in the first row of Table
\ref{table-solutions}. By similar arguments, the
`deconfinement' phase of 5D SYM which coincides with the deconfinement
phase of YM4 must have $W_4 \ne 0$ (for validity of the KK reduction)
and $W_0 \ne 0$ (to exhibit 4D deconfinement). This is shown in the
second row of Table \ref{table-solutions}. 

 The next two rows in Table \ref{table-solutions} describe two
 additional phases of SYM5, which appear only for the (AP,AP)
 b.c. {\em via} the $Z_2$ symmetry.  The third row describes the $Z_2$
 mirror of the confinement phase of YM4, which is characterized by
 $W_0 \neq 0$ and $W_4 = 0$. Since $W_4 = 0$, this phase is a fully 5
 dimensional object and is not related to the YM4 (indeed, so far as
the Polyakov loop order parameters are concerned, this phase can be
identified with a deconfinement phase of SYM5). Another phase is
 the mirror of the deconfinement phase in YM4, which is characterized
 by $W_0 \neq 0$ and $W_4 \neq 0$.  Since this phase has the same
 order parameters as the original deconfinement phase, it is not
 obvious whether these two phases are distinct or smoothly connected
 through a cross over (we represent the possible phase
 boundary/crossover by the horizontal dotted line in Figure
 \ref{fig-D4-AP-phase}).

To describe the next three rows in Table \ref{table-solutions}, note
that the order parameters $(W_0, W_4)$ have a manifestation in
gravity. Thus, e.g. if the $Z_N$ symmetry along $t$ is preserved
($W_0=0$), then, due to the `large $N$ volume independence' mentioned
in Section \ref{sec-QCD}, the free energy $F$ should be independent of
$\b$ (equivalently, the action should be linear in $\b$). This is
satisfied by the action in \eq{action-SD4}. Hence for the parameter
regime in which the solitonic D4 brane is the dominant phase of the
theory, the 5 dim SYM dual exists in a phase characterized by
$W_0=0$. Similarly, the non-linear dependence of the action
\eq{action-SD4} on $L_4$ implies that the gauge theory dual has $W_4
\ne 0$. These results are represented by the 5th row of Table
\ref{table-solutions}. One can similarly deduce by looking at the $\b,
L_4$ dependence of \eq{action-BD4} that for the phase of the gauge
theory represented by the black D4 solution\footnote{The action 
(\ref{action-BD4}) depends linearly on $L_4$, if we rewrite it by using $\lambda_5=\lambda_4 L_4$. Hence the action per unit length in the $x_4$
direction is independent of $L_4$; therefore,  $W_4 =0$.}, 
$W_4 =0, W_0 \ne 0$, 
which could also be inferred by the $t \leftrightarrow x_4$
symmetry. This is represented by the 6th row of Table
\ref{table-solutions}. In the last row, we have included a phase of
SYM5 with (P, AP) fermion b.c. along ($t, x_4$) which corresponds
to a localized D3 soliton, to be discussed in the next section.

We now come back to the conventional proposal of holographic QCD
discussed in the beginning of this section; let us define this
proposal as a combination of three propositions:\\

(a) the solitonic D4 brane phase of gravity corresponds to the
confinement phase of YM4,\\

(b) the black D4 corresponds to the deconfinement phase of YM4, and\\

(c) the SS transition between the two gravity solutions
corresponds to the confinement/deconfinement transition of YM4.\\

In terms of Table \ref{table-solutions}, both the solitonic D4 and the
confinement phase of YM4 (regarded as a phase of SYM) have $W_0 =0,
W_4 \ne 0$; hence both of these phases have the same $Z_N \times Z_N$
symmetry and (a) can in principle be valid. However, (b) cannot be
valid, since the black D4 phase ($W_4=0, W_0 \ne 0$) and the
deconfinement phase of YM4 ($W_4 \ne 0, W_0 \ne 0$) have different
$Z_N \times Z_N$ symmetry. In particular, {\em the black D4
  corresponds to a phase that cannot be KK reduced to YM4 at all
  because of vanishing $W_4$.}
   In fact, it would correspond to 
the $Z_2$ mirror of the confinement phase of YM4 (which 
can be identified with a deconfinement phase of SYM5
which is intrinsically 5 dimensional). 
   Thus, there must be at least one phase boundary
between the deconfinement phase of YM4 and the black D4 phase (see Figure
\ref{fig-D4-AP-phase} for one possibility). Regarding (c), if the SS
transition persists even at weak coupling, then it clearly cannot
coincide with the confinement/deconfinement transition. The reason is
that, even at weak coupling the SS transition can only occur at $\b =
L_4$ (since even stringy effects should satisfy the $Z_2$ symmetry),
whereas it is known that the confinement/deconfinement phase
transition temperature of YM4 goes down ($\b_c$ goes up) at smaller
values of $\l_4$ (see Figure \ref{fig-D4-AP-phase}). 

Through the arguments in this section, it is obvious that the black D4
brane solution does not correspond to the deconfinement phase in the 4
dimensional Yang-Mills theory and the SS transition does not
correspond to the confinement/deconfinement transition.  In the next
section, we propose an alternative correspondence, which resolves
these problems.

\section{Our proposal for the confinement/deconfinement transition in holographic QCD}
\label{sec-phase}

In the previous section, we considered the conventional picture of
holographic QCD at finite temperature and discussed some of the
problems in describing phases of YM4. In this section, we propose an
alternative picture to resolve these problems.

In the above discussion, we described finite temperature gauge theory
by choosing AP b.c. for fermions on the brane along the compactified
Euclidean time.
However, let us recall that the gauge theory
of interest here is pure Yang Mills theory in four dimensions,
which does not have fermions.  Fermions reappear when the validity
condition \eq{gravity-cond-SD4} is enforced and $\L_{YM}$ goes above
the KK scales $1/L_4$. In this sense, fermions are an artifact of the
holographic method and in the region of validity of the pure YM
theory, the periodicity of the fermion should not affect the gauge
theory results.

If we start with the 5 dimensional SYM on $S^1_\beta \times S^1_{L_4} $
with a periodic temporal circle, the partition function is the twisted
one: Tr
$(-1)^F \exp[-\beta H_{\rm SYM5}]$, and  is not the standard
thermal partition function.  However, under the limit $\lambda_4 \ll1$
and $\beta \gg L_4/\lambda_4$, in which the 4 dimensional YM theory
appears, the fermions are effectively decoupled; hence $F \to 0$.  Thus in
this limit the `twisted' partition becomes the usual thermal partition
function, even with the periodic b.c. for the fermions. More
precisely, we have  
\begin{align} 
Z_{\rm (P,AP)}^{{\rm SYM}5}&=\Tr (-1)^F e^{-\beta H_{ {\rm SYM}5}} \nonumber \\
& = 
\Tr e^{-\beta H_{ {\rm YM}4}}+O(e^{-\beta \lambda_4/L_4}), \quad (\lambda_4 \to 0, L_4/\lambda_4 \beta \to 0).
\nn
Z_{\rm (AP,AP)}^{{\rm SYM}5}&=\Tr  e^{-\beta H_{ {\rm SYM}5}} \nonumber \\
& = 
\Tr e^{-\beta H_{ {\rm YM}4}}+O(e^{-\beta \lambda_4/L_4}), \quad (\lambda_4 \to 0, L_4/\lambda_4 \beta \to 0).
\label{partition-function}
\end{align} 
Thus, independent of the boundary condition along the $S^1_\beta$, we obtain the thermal partition function of YM4.

As a consequence, it is pertinent to study the phase structure of the 5
dimensional SYM on $S^1_\beta \times S^1_{L_4} $ with (P,AP) boundary
condition through holography\footnote{The phase structure of the D1
  brane on a $T^2$ with (P,AP) and (AP,AP) has been studied in
  \cite{Aharony:2005ew}. Their results are similar to our D4 brane
  results.  The analysis in \cite{Hanada:2007wn} is also related to
  our study. See also \cite{Mandal:2011hb} for related
comments.}. Indeed, in this sector, we will find a phase transition in gravity,
which continues naturally, at weak coupling, to the confinement/deconfinement transition in
the 4D Yang-Mills theory.
  
\subsection{Gregory-Laflamme transition in the gravity theory with 
(P,AP) boundary condition \label{sec-GL}}

 In this section, we reconsider the gravity theory described in
  Section \ref{sec-problem}, now compactified on $S^1_\b \times S^1_{L_4}$
  with (P,AP) b.c. for fermions. In this case, contrary to the
(AP,AP) case, the black D4 brane solution (\ref{metric-BD4}) is not
allowed (see the comment below \eq{smooth-cond-BD4}).  However a
non-trivial phase boundary still exists, as we will see shortly. The
detailed phase structure for this boundary condition is summarized in
\cite{Harmark:2004ws, Harmark:2007md} (with $t$ and $x_4$ exchanged);
see also \cite{Mandal:2011hb}. We will now outline the salient
features.

Let us start from large $\beta$. In this region the solitonic D4
brane solution (\ref{metric-SD4}) is thermodynamically dominant.  
As we decrease $\beta$, it reaches $O\left(L_4/\sqrt{\lambda_4}\right)$, which is the order of the effective string length at $u=u_0$ \cite{Aharony:1999ti, Gross:1998gk}.
Below this value, winding modes of the string wrapping on the $\beta$-cycle would be excited\footnote{In some previous studies including our works \cite{Mandal:2011hb, Mandal:2009vz}, the winding modes were not correctly evaluated. 
We would like to thank S.~Sugimoto and K.~Hashimoto for discussing this point.}.
Thus the gravity description given
by \eq{metric-SD4} would be valid only if
\begin{align} 
\beta \gg \frac{L_4}{\sqrt{\lambda_4}}.
\label{winding-SD4}
\end{align} 
In order to avoid this problem,
we perform a T-duality along the $t$-cycle and go to the IIB
frame\footnote{\label{ftnt-0B} In the (AP, AP) case, if we
were to perform a T-duality along the $t$-cycle, the theory would
be mapped to the 0B frame which involves a tachyon. The analysis in the bulk gravity in that case would be different from that in the IIB case (we would like to thank S. Minwalla for pointing this out to us). 
 However, unlike in the (P, AP) case, the issue of light winding modes in the 
temperature range \eq{winding-SD4} and the consequent need to go to a T-dual
frame do not arise, since way before we reach temperatures such as
\eq{winding-SD4}, the solitonic D4 solution ceases to be the dominant
solution, and is replaced by the black D4 solution at the temperature
\eq{SS-4}. Recall that in the black brane phase the winding number is not a conserved quantity, as the time circle is contractible; thus there are no winding modes in the spectrum. We hope to come back to a more
detailed analysis of the (AP,AP) case in a future work.}, 
where  the solitonic D4 solution becomes a solitonic D3 brane
uniformly smeared on the dual $t$-cycle.  
The metric describing this solution is given by
\begin{align}
ds^2 =& \a' \left[\frac{u^{3/2}}{\sqrt{d_4 \lambda_5 }}
\left( \sum_{i=1}^{3} 
dx_i^2+f_4(u) dx_4^2 \right)+ \frac{\sqrt{d_4 \lambda_5 }}{u^{3/2}}\left(  \frac{du^2}{f_4(u)} + d{t'}^2
+ u^2 d\Omega_{4}^2 \right)  \right].
\label{metric-SSD3}  
\end{align}
Here $ \alpha' t'$ is the dual of $t$, hence $t'$ has a periodicity $\beta'=(2\pi)^2/\beta=(2\pi)^2T$.  
By considering winding modes on this metric, we can see that the gravity description
is now valid if $\beta \ll L_4/\sqrt{\lambda_4} $.  Thus even when $\beta$ is below
(\ref{winding-SD4}), the gravity analysis in the IIB frame is
possible.

As $\beta$ decreases (hence the dual radius in the IIB frame
increases), the uniformly smeared solitonic D3 brane becomes
thermodynamically unstable at a certain temperature. This temperature,
called the Gregory-Laflamme (GL) instability point, is numerically given by
\cite{Harmark:2004ws}\footnote{
We obtain this result from Eqn. (12.11) in \cite{Harmark:2004ws} with $p=4$ through identification $\hat{L}=\beta$ and $\hat{T}=1/L_4$.}:
\begin{align} 
\beta_{GL,inst} \simeq  7.19 \frac{L_4}{\lambda_4}.
\label{GL-point}
\end{align} 
It is expected that, even before $\beta$ is lowered all the way to
$\b_{GL,inst}$, the smeared solitonic D3 brane becomes meta-stable and
undergoes a first order Gregory-Laflamme (GL) transition at
an inverse temperature $\b_{GL} >
\b_{GL,inst}$,\footnote{The GL instability was initially found in the black
  string. However similar instability exists in the solitonic
  solutions as well; the reason is easily understood in the Euclidian
  space, where there are no differences between the black objects and
  solitonic solutions.}  leading to a more stable, solitonic D3 brane
localized on the dual cycle. This localized solution spontaneously breaks translation symmetry along the temporal direction.
 Although it is difficult to derive the
critical temperature $\beta_{GL}$ precisely, we can approximately
evaluate it as follows \cite{Hanada:2007wn}.  The metric of the
localized solitonic D3 brane is approximately given by that of the solitonic
D3 on $R_{9}\times S^1_{L_4}$ for a sufficiently large radius $\b'$ of the dual
cycle \cite{Harmark:2004ws, Harmark:2002tr},
\begin{align}
ds^2 =& \a' \left[\frac{\tilde{u}^{2}}{\sqrt{d_3 \lambda_5/\beta }}
\left( \sum_{i=1}^{3} 
dx_i^2+f_3(\tilde{u}) dx_4^2 \right)+ \frac{\sqrt{d_3 \lambda_5/\beta }}{\tilde{u}^{2}}\left(  \frac{d\tilde{u}^2}{f_3(\tilde{u})} + \tilde{u}^2 d\Omega_{5}^2  \right)  \right] , \nn
&f_3(\tilde{u})=1-\left( \frac{\tilde{u}_0}{\tilde{u}} \right)^4, \quad \tilde{u}_0= \sqrt{d_3 \lambda_5/\beta}\frac{\pi}{2L_4}, \quad e^{\phi}=\frac{\lambda_5}{2\pi N \beta} .
\label{metric-LSD3}
\end{align}
Here $d_3=2$ (see Eq. \eq{dp}). The value of
$\tilde{u}_0$ is determined by the smoothness condition (as in
\eq{u0-beta}).  $\tilde{u}$ is regarded as a radial component of a
spherical coordinate and is related to the cylindrical coordinates
$(u,t')$ of \eq{metric-SSD3} by roughly $\tilde{u}^2 \sim u^2+t'^2$.
Here we implicitly assume that the soliton is localized at $t'=0$.
This approximation is valid if $\tilde{u}_0$ and $\tilde{u}$ are
sufficiently small such that we can ignore the finite size effect
associated with the dual circle \cite{Harmark:2004ws, Harmark:2002tr}
\footnote{Later on, in \eq{metric-bpsD3}, we will
consider a situation in which $\tilde{u}$ is large. In that
case the finite size of the $t'$-cycle is important and we need to include all the images of the localized soliton on this cycle. }.

The classical action of this solution turns out to be
\begin{align} 
S_{SD3}/N^2V_{3}= -C_3  L_4^{-3}.
\label{action-SD3}
\end{align}  
By comparing this classical action with
(\ref{action-SD4}) \footnote{Note that the classical action is
  invariant under the T-duality, hence \eq{action-SD4} describes both
  the solitonic D4 as well as the smeared solitonic D3.}, we can
estimate that the GL transition will happen around
\begin{align}
\beta_{GL}= \frac{C_3}{C_4} \frac{ L_4}{\lambda_4} \simeq 8.54 \frac{
  L_4}{\lambda_4}.
\label{GL-t}
\end{align}
Note that, near this critical point, $\tilde u_0$ is similar in
magnitude to, but less than, the dual radius \cite{Hanada:2007wn}, so
we expect (\ref{metric-LSD3})-\eq{GL-t} to receive some corrections
from the finite size effect.

\begin{figure}[h]
\begin{center}
\includegraphics[scale=.75]{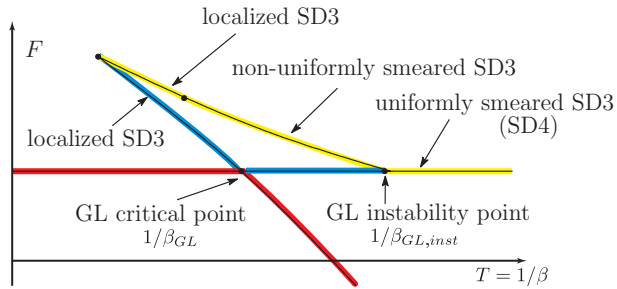}
\caption{The schematic relations among the free energies of various
  solutions in the GL transition (\ref{GL-t}) \cite{Kudoh:2004hs,
    Kol:2004ww}.  The red lines denote the stable solutions.  The blue
  lines denote the meta-stable solutions.  The yellow lines denote the
  unstable solutions.  The solitonic D4 (uniformly smeared solitonic
  D3) becomes unstable at the GL instability point (\ref{GL-point}).
  In addition, the localized solitonic D3 cannot exist below a certain
  temperature.  }
\label{fig-potential}
\end{center}
\end{figure}

In addition to these two solutions, \eq{metric-SSD3} and \eq{metric-LSD3}, 
there is another solution:
solitonic D3 brane non-uniformly smeared on the dual cycle.  The
metric of this solution is perturbatively derived in
\cite{Harmark:2004ws}.  This non-uniform solution arises at the GL instability
point and is always unstable.  Although the complete behaviour of the
non-uniform solution has not been explored, it is expected that it
merges with the localized solitonic D3 as shown in
Figure \ref{fig-potential} \cite{Aharony:2004ig, Harmark:2007md, Kudoh:2004hs, Kol:2004ww}\footnote{In this paper, 
we will often use the notation ``SD$p$'' for solitonic D$p$ brane(s).}.  
The fact that the GL  instability temperature
(\ref{GL-point}) is higher than the approximately obtained GL critical
temperature (\ref{GL-t}) is consistent with this expectation.

Let us make a few comments on the localized solitonic D3 solution
(\ref{metric-LSD3}). Firstly, through a calculation similar to
(\ref{gravity-cond-SD4}), the gravity description can be shown to be
valid (i.e. stringy effects can be ignored) if $\lambda_4 \gg
\beta/L_4$.  
Secondly, this
solution ceases to exist if $\beta$ is too large, as shown in
Figure \ref{fig-potential}; intuitively, if $\beta$ were too large, the
dual cycle would become smaller than the size of the localized soliton, which is not possible. Thirdly, the free energy
$F=S_{SD3}/\beta$ of this solution, unlike that of \eq{metric-SD4}, is
proportional to temperature; hence the entropy of this solution is
non-zero at $O(N^2)$ \footnote{The entropy derived from the classical
  action (\ref{action-SD3}) does not depend on temperature.  If we
  consider higher order corrections of $u_0$ and $u$ in the metric
  (\ref{metric-LSD3}) as in \cite{Harmark:2004ws}, the corrected
  entropy depends on the temperature.} and the temporal $Z_N$ symmetry
is broken.  In addition, because of the non-trivial $L_4$-dependence 
of the classical action \eq{action-SD3}  we can see that the $Z_N$ symmetry along the
$x_4$-cycle is also broken.  Thus the Polyakov loop $W_0 $
and $W_4$ in this solution are both non-zero as shown in Table
\ref{table-solutions} and, contrary to the black D4 solution, the
localized solitonic D3 solution is appropriate for a description of the
deconfinement phase in the dual gauge theory (which also
has $W_0 \ne 0, W_4 \ne 0$). We will build on this observation in
the next subsection.

\subsection{Gregory-Laflamme transition as a confinement/deconfinement transition}
\label{sec-GL-CD}

The phase structure\footnote{In this phase diagram, we do not
  distinguish the solitonic D4 and uniformly smeared solitonic D3
  because they are essentially the same object, although the regions
  of validity of their gravity descriptions are different. } of the 5
dimensional SYM on $S^1_\beta \times S^1_{L_4} $ with the (P,AP)
boundary condition is shown in Figure \ref{fig-D4-phase-P}.  The
strong coupling region (the blue region in Figure
\ref{fig-D4-phase-P}) described by type II supergravity and is
characterized by the GL phase transition which occurs at a temperature
given by (\ref{GL-t}).  In the weak coupling region (the green region
in Figure \ref{fig-D4-phase-P}), the 4 dimensional Yang-Mills is
realized at low temperatures ($\beta \gg L_4/\lambda_4$).  Although
this region is common to the (P,AP) phase diagram in Figure
\ref{fig-D4-phase-P} and the (AP,AP) 
phase diagram in Figure
\ref{fig-D4-AP-phase}, the mirror of this region under the $Z_2$
($\beta \leftrightarrow L_4$) does not exist in Figure
\ref{fig-D4-phase-P} since the (P,AP) b.c is not $Z_2$-symmetric. 

\begin{figure}
\begin{center}
\includegraphics[scale=.75]{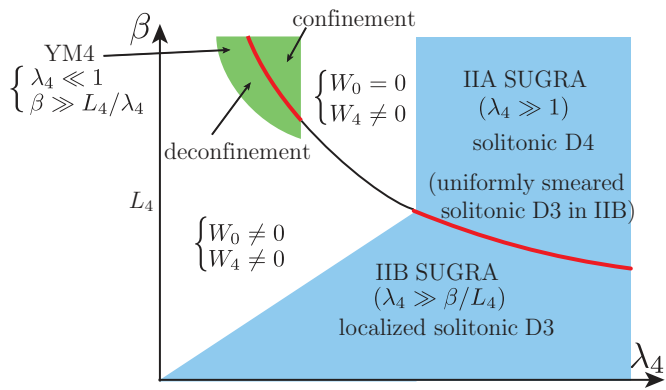}
\caption{The phase structure of five dimensional SYM on $S^1_\beta
  \times S^1_{L_4} $ with the (P,AP) boundary condition.  The gravity
  analysis is valid in the strong coupling region (the blue region).
  The 4 dimensional YM description is valid in the green region.  The
  phase structure in the intermediate region, including the phase
  boundary shown by the solid black line, is an extrapolation from
  these two descriptions.  }
\label{fig-D4-phase-P}
\end{center}
\end{figure}

The main point to emphasize here it that: contrary to the previous
phase structure in the (AP,AP) case, now the localized solitonic D3
phase has the same order parameters ($W_0 \ne 0, W_4 \ne 0$) as the
deconfinement phase, thus making it plausible that these two phases
are smoothly connected.  In Figure \ref{fig-D4-phase-P}, we have
indicated this by assuming the simplest extrapolation through the
region of the intermediate coupling.

\paragraph{Our proposal}

In view of the above observations, we propose a strong coupling
continuation of weakly coupled 4-dimensional Yang Mills theory as
shown in Table \ref{table-proposal}.

\begin{table}
\hspace{-4ex}
{\small
\begin{tabular}{c|c|c| c}
\hline
& Weak coupling  & Strong coupling 
& $Z_N \times Z_N$ 
\\
\hline
 &  4 dimensional YM  &  Gravity (P,AP)& \\
\hline
\hline
Low temp.  &  Confinement phase 
&  Solitonic D4 phase & $W_0=0, W_4 \ne 0$
\\
\hline
High temp. &  Deconfinement phase 
&  Localized solitonic D3  & $W_0 \ne 0, W_4 \ne 0$
\\
\hline
Transition & Confinement/Deconfinement & Gregory-Laflamme &
\\
\hline
\end{tabular}}
\caption{Our proposal.} 
\label{table-proposal}
\end{table}

Of course, our proposal is based on a simple extrapolation between the
intermediate coupling regime, and the real story there could be more
involved.  However the mere existence of such a simple extrapolation
is a significant improvement over the previous proposal in the (AP,AP)
case, where we are certain that there has to be at least one phase
boundary between the deconfinement phase and black D4 solution (for
the simple reason that their order parameters have different values).
It is clearly important, therefore, to further investigate the nature
of the deconfinement phase in terms of the localized
solitonic D3 solution based on the correspondence outlined above.

\section{New correspondences in holographic QCD}
\label{sec-new}

Using our proposed correspondence, we can explain anew several
phenomena in the gauge theory from gravity.  In this section, we list
some of these phenomena.

\subsection{Polyakov loop and D3 brane distribution}
\label{sec-ZN}

In our proposal, we identified the GL transition in the IIB frame to the
confinement/deconfinement transition in the gauge theory. In the GL
transition, the distribution of the D3 branes on the dual circle
changes from a uniform distribution at low temperatures to a localized
one at high temperatures. In this subsection, we explain what the
corresponding phenomenon is in the gauge theory.

Since the original five-dimensional gauge theory (of the D4 branes)
appears in the IIA frame, in order to understand the role of the D3
branes, we need to consider a T-duality along the temporal direction.
Under this T-duality, $X_0$, which is transverse to the D3 brane, is
mapped to the gauge potential $A_0$ on the D4 brane. Thus the D3 brane
distribution on the dual circle is related to a distribution of
$\theta_k$'s, which are defined as the eigenvalues of the Polyakov
loop operator $ \exp(i \theta_k) \delta_{kl} = \left[P \exp(i\int_0^\beta
A_0 dt)\right]_{kl},$ $k,l=1,...,N$.

If the D3 branes are uniformly distributed, the $\theta_k$ are also
uniformly distributed. In that case, 
by choosing an appropriate gauge, we can take $\theta_k=2\pi k/N$.
Then the temporal Polyakov loop operator (\ref{Polyakov}) becomes 
\begin{align} 
 W_0= \frac1N {\rm Tr } P e^{i \int_0^{\beta} A_0 dx^0 } = \frac{1}{N} \sum_{k=1}^N e^{2\pi i k/N } =0.
\end{align} 
Hence this distribution corresponds to the confinement phase.
  On the other hand, if the D3 branes are localized, then $\theta_k$ are also
localized and the Polyakov loop becomes non-zero, which characterizes the
deconfinement phase.  This observation is consistent with the entropy
arguments in section \ref{sec-GL}.

The above discussion shows a direct relation between the D3 brane
distribution and the eigenvalue distribution of the Polyakov loop
operator. This latter quantity can sometimes be explicitly evaluated
\cite{Mandal:2011hb, Sundborg:1999ue, Aharony:2003sx, Aharony:2005bq, Mandal:2009vz, Morita:2010vi}.
E.g. in \cite{Mandal:2011hb}, for four dimensional Yang Mills theory
on a small $T^2$, we found that the uniform, non-uniform and localized
distribution of $\theta_i$, all appear, with obvious correspondence to
similar gravitational solutions.  Especially the free energies of
these three solutions show a ``swallow tail'' relation similar to
Figure \ref{fig-potential}.  (See Figure 3 in \cite{Mandal:2011hb}.)
This correspondence strongly supports our proposal.

\subsection{Gregory-Laflamme transition as a Hagedorn transition}

Our proposal opens up the interesting possibility of a relation
between the GL transition and the Hagedorn transition.

It is known that the GL instability is an instability of the KK modes of the graviton along the compact circle \cite{Gregory:1994bj}.
In our case, the KK modes along the dual
temporal circle, which are associated with the GL instability at (\ref{GL-point}) in the IIB
description, are mapped to winding modes around the temporal circle
through the T-duality \cite{Harmark:2007md, Ross:2005vh}. 
This then indicates that the type IIB GL transition is associated with the excitation of the winding modes of the IIA string\footnote{Note that we have argued in \ref{sec-GL}  that at values of $\beta$ lower than \eq{winding-SD4}, winding modes get excited. 
On the other hand, the instabilities from the winding modes arise at the much smaller $\beta$ (\ref{GL-point}) for a large $\lambda$.
It would be interesting to understand the mechanism for the separation of these two temperatures in the IIA string theory, although it may be difficult since we need to quantize the strings to investigate it.}.
This phenomenon is similar to the Hagedorn transition in string theory
\cite{Atick:1988si}, where the associated  instability
is caused by temporal winding modes.  
Thus the GL transition in the IIB description might
correspond to the Hagedorn transition in the IIA description.

Note that from large $N$ gauge-theoretic calculations also, 
the usual confinement/deconfinement transition is believed to be related to the Hagedorn transition.
It has been explicitly shown in weakly coupled gauge theories \cite{Sundborg:1999ue, Aharony:2003sx}.  
This makes it plausible that the Hagedorn transition in the Yang-Mills
theory continues to the Hagedorn transition in the IIA string, which,
as we argued above, is possibly the dual of the GL transition in the
IIB supergravity.

\section{\label{sec-chiral} Chiral symmetry restoration 
in Sakai-Sugimoto model}

In the previous sections, we have seen that the conventional
holographic representation of the confinement/deconfinement transition
as the SS transition is fraught with problems, to circumvent which we
proposed in Section \ref{sec-GL-CD} a different interpretation in
terms of a GL transition.  However, the SS transition has widely been
employed in holographic QCD and purports to explain several phenomena
in real QCD.  In particular, the chiral symmetry restoration in the
Sakai-Sugimoto model was neatly explained in \cite{Aharony:2006da}.
In their scenario the black D4 brane plays a crucial role. However,
since in our proposal the black branes do not appear anymore, we need
to find an alternative idea for realizing chiral symmetry restoration.
In this section, we discuss how chiral symmetry restoration can happen
in the localized solitonic D3 background.

\subsection{Sakai-Sugimoto model and chiral symmetry breaking}
\label{sec-Sakai-Sugimoto}

The Sakai-Sugimoto model \cite{Sakai:2004cn} was proposed to describe
low energy hadron physics in holographic QCD and elegantly reproduces
many aspects of the real QCD.  We first briefly review this model and
show how chiral symmetry breaking at low temperatures is realized
in terms of the dual gravity.

The Sakai-Sugimoto model is an extension of the holographic model of
QCD discussed in section \ref{sec-holographicQCD}.  Sakai and Sugimoto
added, to the $N$ D4 brane system, $N_f$ D8 and $\overline{\rm D8}$
branes which are localized on the $x^4$ circle and fill all other
directions as follows 
\begin{eqnarray}
\begin{array}{ccccccccccc}
& (0) & 1 & 2 & 3 & (4) & 5 & 6 & 7 & 8 & 9 \\
{\rm D4} & - & - & - & - & - &&&&& \\
{\rm D8/\overline{D8}}
& - & - & - & - &  & - & - & - &- & - 
\end{array}
\end{eqnarray}
Here $()$ denotes the compactified directions.  (See Figure
\ref{fig-SSmodel} (a).)  This model has a free parameter $L$, which is
the asymptotic ($u \to \infty$) distance between the D8 and
$\overline{\rm D8}$ brane on the $x_4$ circle.  In the original
Sakai-Sugimoto model \cite{Sakai:2004cn} $L$ was taken to be
  $L_4/2$; however, we will let it be general and take values in $[0,
    L_4/2]$ (the other half, $L \in (L_4/2, L_4)$ is related by a
  reflection and need not be considered separately).
  
This model has $U(N_f) \times U(N_f)$ gauge symmetry on the $N_f$ D8 and
$\overline{\rm D8}$ branes, which can be interpreted as a chiral
$U(N_f)_L \times U(N_f)_R$ flavor symmetry in QCD.  As we will see
soon, in some situations, the D8 and $\overline{\rm D8}$ branes merge
and the chiral symmetry is broken to a single $U(N_f)$.  Sakai and
Sugimoto proposed that this is the holographic realization of chiral
symmetry breaking \cite{Sakai:2004cn}.

Let us take a large $N$ limit {\em a la} Maldacena and, according to
the principle of holography, replace the $N$ D4 branes with a
corresponding gravity solution.  Here we will keep $N_f \ll N$ such
that we can ignore the back-reaction of the D8/$\overline{\rm D8}$
branes onto the background geometry (this is the so-called `probe
  approximation').  In that case, the background (D4 brane) geometry
  is determined thermodynamically (as the dominant classical solution
  at a given temperature, as we have done in sections
  \ref{sec-problem} and \ref{sec-phase}), and the D8/$\overline{\rm
    D8}$ brane configuration coupled to this background, is determined
  dynamically, with the given distance $L$ as a boundary
  condition.

Chiral symmetry breaking in this model happens as follows. At
sufficiently low temperatures, the favoured geometric background is
that of solitonic D4 solution (\ref{metric-SD4}), as we have seen in
section \ref{sec-problem} and \ref{sec-phase}.  On this background, the D8 and
$\overline{\rm D8}$ cannot extend separately and need to merge as
shown in Figure \ref{fig-SSmodel} (b).  As a
result, the $U(N_f) \times U(N_f)$ gauge symmetry on the D8 and
$\overline{\rm D8}$ is broken to $U(N_f)$, representing chiral
symmetry breaking ($\chi$SB) in the dual gauge theory.

\begin{figure}
\begin{center}
\includegraphics[scale=.50]{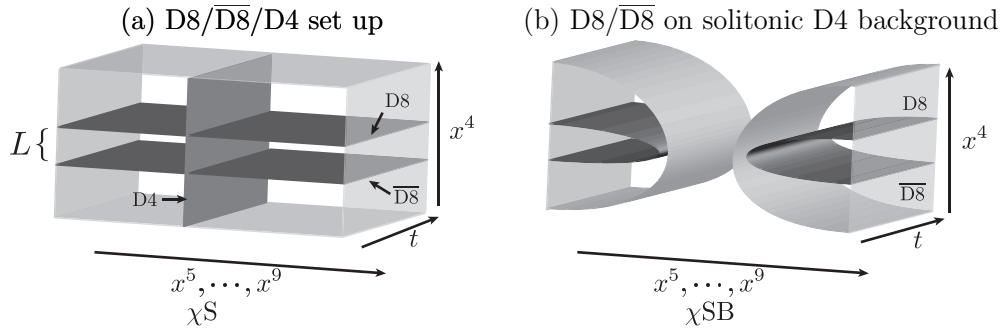}
\caption{Brane configuration of the Sakai-Sugimoto model at a finite
  temperature. Figure (a) shows the basic brane configuration of the
  Sakai-Sugimoto model before considering the large $N$ limit. The
  long horizontal direction schematically denotes the $x^5,\cdots,x^9$
  coordinates, which are transverse to the D4 branes (if the D4 branes
  are localized at $x^5=\cdots =x^9=0$ in Cartesian coordinates, the
  horizontal line can be taken as, for example, $x^5$ with $x^6=\cdots
  =x^9=0$). The directions $x^1, x^2, x^3$ are common to the D4 and
  D8/$\overline{\rm D8}$ branes and are not shown. The directions $t$
  and $x^4$ are periodic and we should identify `up' and `down', and
  `front' and `back' planes in the figure, respectively. The square
  planes on the edges denote $T^2=S^1_t \times S^1_{x^4}$ at
  $u=\infty$ ($x^5=\pm \infty$).  The $N_f$ D8 and $\overline{\rm D8}$
  branes are localized on $S^1_{x^4}$ with the asymptotic ( $u\to
    \infty)$ distance $L$.  The world-volumes of the $N$ D4
  branes are depicted by vertical planes and those of the D8 and
  $\overline{\rm D8}$ by horizontal planes. Before taking the large
  $N$ limit, the D8 and anti-D8 branes extend parallelly and chiral
  symmetry is preserved.  In (b) we consider the large $N$ limit and
  replace the $N$ D4 branes by the solitonic D4 brane geometry
  (\ref{metric-SD4}).  This geometry is pinched off at $u=u_0$ and D8
  and $\overline{\rm D8}$ cannot extend to the $x^5,\cdots,x^9$
  direction.  Thus they need to merge and chiral symmetry is broken.}
\label{fig-SSmodel}
\end{center}
\end{figure}

\subsection{Chiral symmetry restoration in the black D4 brane background}

In the gauge theory, it is expected that chiral symmetry is restored
at a sufficiently high temperature.  Thus if holographic QCD is to
work, there should be a corresponding phenomena in the dual gravity
description.

In \cite{Aharony:2006da}, a mechanism for chiral symmetry restoration
was suggested by considering D8 and $\overline{\rm D8}$ in the black
D4 brane background (\ref{metric-BD4}) in the (AP,AP) case.  
Contrary to the solitonic D4 brane geometry
(\ref{metric-SD4}), the $u-t$ plane of the black D4 background has a `cigar' geometry. Thus the D8 and
$\overline{\rm D8}$ branes can wrap the cigar separately as shown in
Figure \ref{fig-SSmodel-SS} (c). If such a configuration is
energetically favoured, $U(N_f) \times U(N_f)$ gauge symmetry is
preserved and chiral symmetry is restored.

\begin{figure}
\begin{center}
\includegraphics[scale=.50]{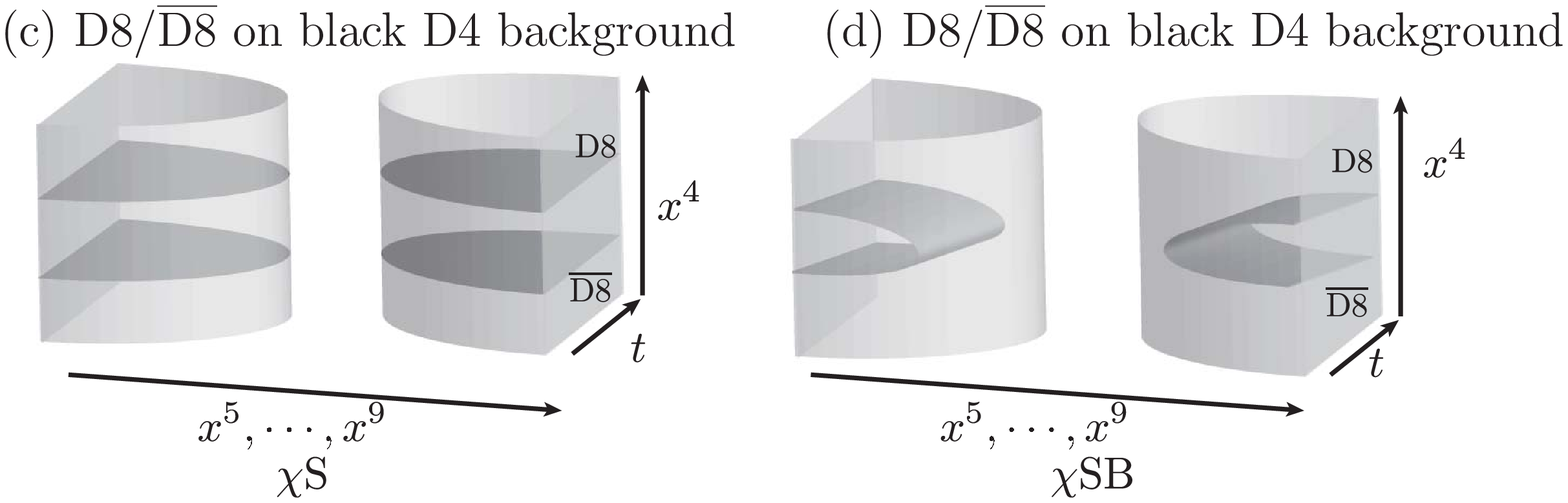}
\caption{Chiral symmetry restoration in the black D4 brane background (\ref{metric-BD4})
  \cite{Aharony:2006da}. In this background, although the horizontal
  direction is still pinched off, the $t-u$ plane has a `cigar'
  geometry and the D8 and $\overline{\rm D8}$ branes can separately
  wrap the cigar as shown in (c).  (Recall that the `front' and the
  `back' of this figure are identical.)  In this case, chiral symmetry
  is restored.  On the other hand, the configuration shown in (d) is
  also possible.  This configuration is similar to (b) and chiral
  symmetry is broken.}
\label{fig-SSmodel-SS}
\end{center}
\end{figure}

In addition to this configuration, another configuration shown in
Figure \ref{fig-SSmodel-SS} (d) is possible in the black D4
background.  This configuration is similar to Figure
\ref{fig-SSmodel-SS} (b) and chiral symmetry is broken.  
In \cite{Aharony:2006da}, energies of these two configurations are
  compared with the black D4 background. It was found (see Figure 7 of
  \cite{Aharony:2006da}) that for $L > L^{\rm SS}_c \equiv 0.154 L_4$,
  it is configuration (c) which is always favoured, indicating a
  concurrence of deconfinement and chiral symmetry restoration
  transitions at $T_{SS}= 1/\b_{SS} \equiv 1/L_4$, while for $L<
  L^{\rm SS}_c$, a new window $T_{SS} \le T < 0.154/L$ opens up where
configuration (d) is the favoured one, indicating coexistence of deconfinement and broken chiral symmetry.

\subsection{Chiral symmetry restoration in the localized solitonic D3 brane.}
\label{sec-chiral-SD3}

Although the chiral symmetry restoration was explained, as above, in
the black D4 brane background, our main thesis in this paper is that
the black D4 solution itself is fraught with problems, if we interpret
this solution as the deconfinement phase of the four dimensional gauge
theory (see Section \ref{sec-AP}).  The alternative we proposed in
Section \ref{sec-GL-CD} is that the localized solitonic D3 brane
solution (\ref{metric-LSD3}) in the (P,AP) b.c. should be taken as the correct
representation of the deconfinement phase in the dual gauge theory.
In view of this, we need to understand chiral symmetry restoration in
this solution instead of in the black D4 solution.

One subtlety in the application of our proposal to the Sakai-Sugimoto model is the existence of the fundamental quarks from the open string between the D4 and D8/$\overline{\rm D8}$ brane. 
These quarks are, of course, not decoupled in the 4 dimensional limit $\lambda_4,L_4/\beta \lambda_4 \to 0$.
Hence, unlike in case of the adjoint fermions, 
where the P and the AP b.c. on the thermal cycle reduce
to the same quantity in the four dimensional limit (see \eq{partition-function}),
for the fundamental quarks the P and the AP b.c's {\em  differ in the 4D limit}; hence, we must choose the AP b.c. on the temporal cycle to investigate 
the thermodynamics of the Sakai-Sugimoto model. Since our
proposal crucially uses P b.c. for the fermions along the temporal cycle,
we must address this issue \footnote{\label{nf-vs-n}Of course, for
$N \gg N_f$, the fractional difference in the free energy
between the P b.c. and the AP b.c., which entirely arises
from the fundamental quarks in the 4D limit, is $O(N_f/N) 
\to 0$. However, such fractional differences are
important when we want to discuss the free energy of the quarks,
which  itself is of order $O(N_f/N)$ compared to that of
the gluons.}.
It turns out that with a small addition to our model, {\em viz.}
that of an imaginary chemical potential \cite{PRINT-86-0360 (BRITISH-COLUMBIA)}, we can obtain the desired AP b.c. around the temporal cycle  for the quarks {\em retaining the P b.c. for the adjoint fermions}\footnote{We  would like to thank Y. Hidaka for suggesting this point.}.
Details of this approach will appear in a forthcoming paper 
\cite{MM-progress}. Introduction of such a chemical
potential does not cause any
essential modification to the analysis
that follows below, as shown in detail in \cite{MM-progress}.  
Consequently, we will proceed below with periodic (fundamental as well
as adjoint) fermions. 

\begin{figure}
\begin{center}
\includegraphics[scale=.50]{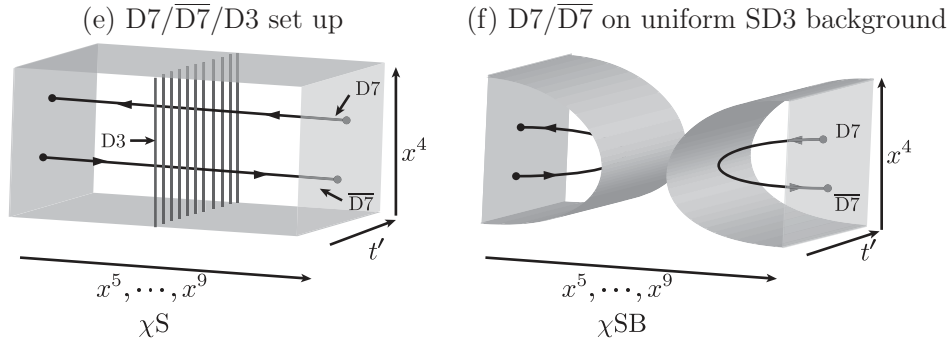}
\caption{T-dual of Figure \ref{fig-SSmodel} along the $t$-cycle.  (e)
  and (f) are the T-duals of (a) and (b) respectively.  The
  distribution of the D3 branes (schematically represented by the
  bunch of vertical lines) on $S^1_{t'}$ is fixed dynamically.}
\label{fig-SSmodel-GL}
\end{center}
\end{figure}

In our proposal, the gravity analysis has been done in IIB supergravity by performing a T-duality along the $t$-cycle: $S^1_t \to S^1_{t'}$. Thus it is convenient to dualize the
brane configuration of the Sakai-Sugimoto model to the IIB frame.
Since all the D branes in the Sakai-Sugimoto model wrap the $S^1_t$,
the above T-duality maps the D4 branes and D8/$\overline{\rm D8}$
branes to D3 and D7/$\overline{\rm D7}$ branes spreading
\begin{eqnarray}
\begin{array}{ccccccccccc}
& (0') & 1 & 2 & 3 & (4) & 5 & 6 & 7 & 8 & 9 \\
{\rm D3} &  & - & - & - & - &&&&& \\
{\rm D7/\overline{D7}}
&  & - & - & - &  & - & - & - &- & - 
\end{array}
\end{eqnarray}
See Figure \ref{fig-SSmodel-GL} (e) also. The distribution along $S^1_{t'}$ of the D
branes in the IIB description (related to the gauge field $A_0$ in the
IIA description), is determined dynamically.  In the probe
approximation, according to the analyses in section \ref{sec-GL},
at temperatures below the GL critical temperature (\ref{GL-t}), the D3
branes are distributed uniformly and, above it, they are localized on
the $S^1_{t'}$. The fate of the chiral gauge group in the gravity
representation depends on the stable configurations of the probe
D7/$\overline{\rm D7}$ branes in these backgrounds.

In the uniformly smeared solitonic D3 brane geometry (\ref{metric-SSD3}) shown in Figure
\ref{fig-SSmodel-GL} (f), the situation is similar to (b) in Figure
\ref{fig-SSmodel}.  Since the $u$ direction is smoothly pinched off at
$u_0$, the D7 and $\overline{\rm D7}$ have to merge and chiral
symmetry is broken.

On the other hand, in the localized solitonic D3 brane geometry, the
horizontal direction is not fully pinched off.  
Recall that the
geometry (\ref{metric-LSD3}) is pinched off at
$\tilde{u}=\tilde{u}_0$, where $\tilde{u} \sim \sqrt{u^2+t'^2}$ and $u^2=(x_5^2+...+x_9^2)/\alpha'^2$.
 Thus $u$ can reach zero.  As a result, D7 and $\overline{\rm D7}$ can
extend separately as shown in Figure \ref{fig-SSmodel-GL2} (g).  This
configuration is similar to (e) in Figure \ref{fig-SSmodel-GL} and
would restore chiral symmetry.  In addition to this configuration, a
chiral symmetry broken configuration is also possible as shown in
Figure \ref{fig-SSmodel-GL2} (h).

\begin{figure}
\begin{center}
\includegraphics[scale=.50]{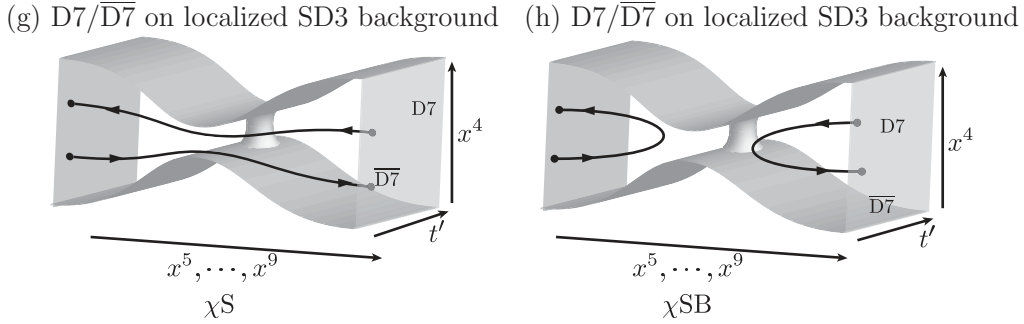}
\caption{Chiral symmetry restoration in the localized smeared
  solitonic D3 brane background.  In this geometry, the D3 branes are
  localized on the $S^1_{t'}$ and the horizontal $x^5,\cdots,x^9$
  direction is not fully pinched off.  Then we can consider two
  configurations.  The chiral symmetry is preserved in (g) and broken
  in (h).}
\label{fig-SSmodel-GL2}
\end{center}
\end{figure}

The last task is the evaluation of the stability of these two
configurations at temperatures higher than $T_{GL}=1/\beta_{GL}$ (\ref{GL-t}).  In the flat
space, the force between a single D3 and single D7 (or between a
single D3 and a single $\overline{\rm D7}$) in our configuration is
repulsive, since the number of the Neumann-Dirichlet open strings
between them is 6 \cite{Polchinski:1998rr}.  
Thus we expect that even after taking the
large $N$ and near horizon limits, the force may be repulsive.  As a
result, the D7/$\overline{\rm D7}$ branes, which are separated from
the localized SD3 in the $t'$-direction, would try to move away as far
as possible in this direction. However, since the $t'$-direction is
compactified on a circle, the D7/$\overline{\rm D7}$ branes should
then end as being fixed at the point on the $t'$-circle which is
antipodal to the localized SD3.  E.g. if we put the localized SD3
brane at $t'=0$, the D7 as well as the $\overline{\rm D7}$ branes will both be at
$t'=\beta'/2$. See the left diagram of  Figure \ref{fig-probeD7}.
In this case, we can effectively restrict the dynamics of these branes to ($u-x_4$) plane (given by $t'=\b'/2$).
Thus the problem reduces to finding classical solutions $x_4(u)$ with a boundary condition: $x_4(u)
\to L/2$ ($u \to \infty$), where we put the D7 at $x_4=
L/2$ and $\overline{\rm D7}$ at $-L/2$.

\begin{figure}
\begin{center}
\includegraphics[scale=.60]{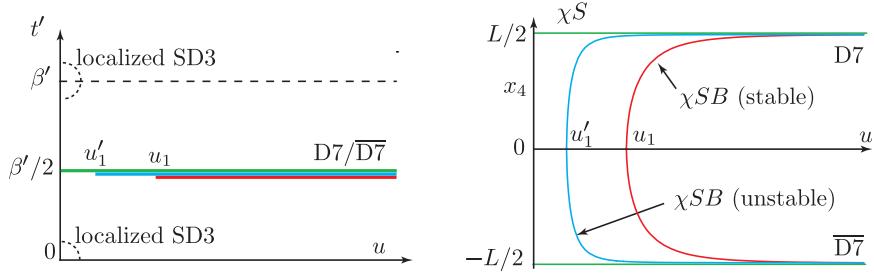}
\caption{Three classical configurations of the probe
  D7/$\overline{{\rm D7}}$ brane on the localized SD3 background in
  $u-t'$ and $u-x_4$ space.  Green lines correspond to the
  configuration of the D7 and $\overline{{\rm D7}}$ branes in Figure
  \ref{fig-SSmodel-GL2}(g), which preserve chiral symmetry.  The red
  and blue lines correspond to Figure \ref{fig-SSmodel-GL2}(h), which
  break chiral symmetry.  These three solutions satisfy the same
  boundary condition $x_4 \to L/2$ ($u \to \infty$).  The two $\chi
  SB$ solutions appear only for $L<L_*(T)$.  The energy of these
  solutions are plotted in Figure \ref{fig-fene} (a).  }
\label{fig-probeD7}
\end{center}
\end{figure}

The problem stated above is difficult to solve precisely near the GL
transition since the background metric (\ref{metric-LSD3}) is only an
approximate description. The metric around the D7/$\overline{{\rm
    D7}}$ brane (at $t'=\beta'/2$) becomes more and more accurate,
however, when $T \gg T_{GL}$.  As a result, we solve for the stable
configuration in this limit. Details of the calculation are presented
in appendix \ref{app-probe}.  We find three solutions, corresponding
to the D brane configurations of Figures \ref{fig-SSmodel-GL2} (g) and
(h), as shown in Figure \ref{fig-probeD7}.  In the appendix, we
compare the classical DBI actions of these solutions numerically, as
depicted in Figure \ref{fig-fene} (a) (see also the phase diagram in
Figure \ref{fig-tc}): as one increases temperature beyond a certain
value, chiral symmetry is restored, the transition being of first
order.  Therefore, one can see that even in our proposal, similarly to
the case of the black D4 \cite{Aharony:2006da}, we can explain chiral
symmetry restoration at high enough temperatures. 

\begin{figure}
\begin{center}
\includegraphics[scale=0.9]{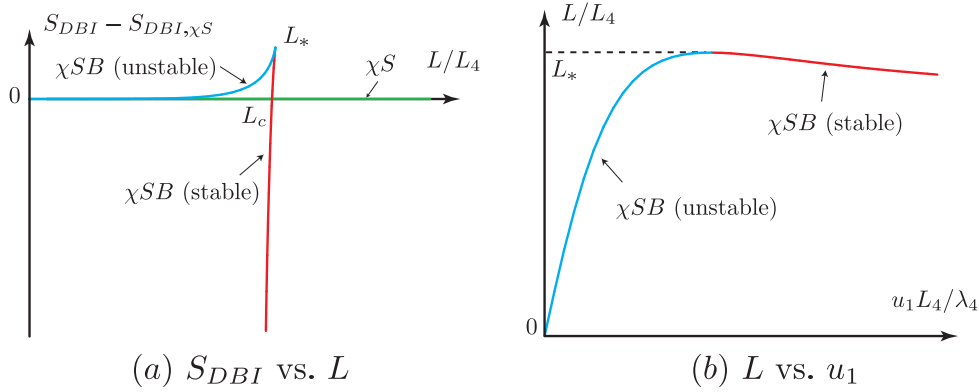}
\caption{ (a) The DBI action for probe D7 branes vs.~$L$.  We
  evaluated (\ref{action-DBI}) for the three solutions (one $\chi S$
  and two $\chi SB$ solutions).  Since these DBI actions diverge, we
  regularize them by subtracting the DBI action for the $\chi S$
  solution.  A first order transition occurs at $L=L_c$.  (b) The
  relation between $u_1$ and $L$ in the $\chi SB$ solution
  (\ref{x_4}).  Two $\chi SB$ solutions exist for a given $L$ if $L <
  L_*$.  The colour coding in these figures is the same as that in
  Figure \ref{fig-probeD7}.}  
\label{fig-fene}
\end{center}
\end{figure}
\begin{figure}
\begin{center}
\includegraphics[scale=.9]{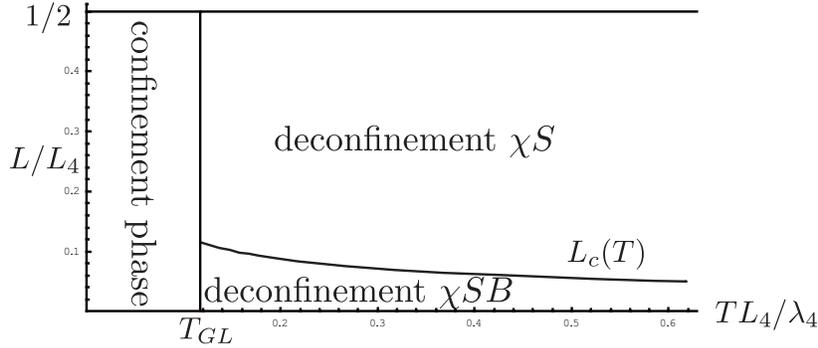}
\caption{Preliminary phase diagram of the Sakai-Sugimoto model.  We
  numerically evaluate $L_c(T)$ by using the high temperature
  approximation and the result near $T_{GL}$ is not reliable.  The $T$
  coordinate is scaled by $\lambda_4/L_4$.  In higher temperature, the
  $\chi S$ phase is favoured.}
\label{fig-tc}
\end{center}
\end{figure}

\section{\label{Conclusions}Conclusions}

In this paper, we showed that the conventional representation of the
confinement/deconfinement transition in holographic QCD has several
problems and proposed an alternative representation which resolves
these problems.

As mentioned earlier, problems similar to the above had also been
encountered in the study of two dimensional bosonic gauge theory in
\cite{Mandal:2011hb}. 
This indicates that the issues addressed in this paper are rather general in the discussion of holography for non-supersymmetric gauge theories at finite temperatures. 
To elaborate, in the standard holographic
procedure, a $p$-dimensional non-supersymmetric gauge theory is first
constructed through the KK reduction of a $(p+1)$-dimensional super
Yang-Mills theory on a Scherk-Schwarz circle.  The $(p+1)$-dimensional
SYM at large $N$ can be mapped to a scaling limit of D$p$ brane
geometries \cite{Itzhaki:1998dd}. At finite temperatures, because of
the two compact cycles (temporal and Scherk-Schwarz), several distinct
solutions (depending on boundary conditions) appear in gravity as
shown in section  \ref{sec-problem} and \ref{sec-phase}: solitonic D$p$ (equivalently,
uniformly smeared solitonic D$(p-1)$), localized solitonic D$(p-1)$,
and black D$p$.  The black D$p$ brane solution appears at high temperatures in the (AP,AP) case, while
the localized solitonic D$(p-1)$ brane solution is the high temperature phase in
the (P,AP) case. A table similar to Table \ref{table-solutions} can
again be constructed, where the appropriate order parameters would be
$W_{p}, W_0$; these would again appear to favour the localized solitonic
D$(p-1)$ phase as the more suitable representation of the
deconfinement of $p$-dimensional YM theory (rather than the more
conventional black D$p$ phase which appears only in the (AP,AP) b.c.).
Following this logic, a $p$-dimensional analogue of our proposal (see
Section \ref{sec-GL-CD}) would appear to give a better description of
holographic QCD in $p$ dimensions. In particular, we believe that the
Gregory-Laflamme transition between the solitonic D$p$ and localized
solitonic D$(p-1)$, would, as in this paper, be related to the
confinement/deconfinement transition in the $p$-dimensional gauge
theory.

\subsection{Further questions}

In order to further understand holographic QCD through the above
proposal, it would be of interest to address the following questions.

\paragraph{Transition temperature}

From (\ref{GL-t}), the critical temperature of the
confinement/deconfinement transition (the GL transition) would be
$O(\lambda_4/L_4)$.  On the other hand, holographic QCD seems to predict that the square root of the QCD string tension is
$O(\sqrt{\lambda_4}/L_4)$, as in (\ref{winding-SD4}), whereas the glueball masses are $O(1/L_4)$
\cite{Aharony:1999ti}.
 It would be important to understand the reason
for the separation of these scales and how they evolve from strong coupling to weak coupling.

\paragraph{Quantitative correspondence}

Although our new proposal reproduces the known qualitative features of
4 dimensional Yang-Mills theory, it does not automatically lead to a
quantitative agreement.  For example, the free energy of the
deconfinement phase of the YM theory at a sufficiently high
temperature must be proportional to $T^4$, since the coupling becomes
weak and the theory becomes approximately conformal.
However the free energy of the localized solitonic D3 is proportional
to $T$ (as can be seen from \eq{action-SD3}). 
This is not entirely surprising since the functional form of $F(T)$ can change as one evolves from weak coupling to strong coupling. Furthermore, one has to exercise caution in defining a high temperature limit of YM4 in the holographic context since the temperature must
always remain much smaller than the KK scale.

\paragraph{YM4 from SYM5 with (AP,AP)}
In this article, we have emphasized the correspondence between YM4
and SYM5 with (P,AP) b.c.  However, as we pointed out in 
\eq{partition-function}, the 5 dimensional SYM with (AP,AP)
b.c. should also be related to YM4, since the boundary condition
becomes irrelevant in the limit $\lambda_4 \ll1$ and $\beta \gg
L_4/\lambda_4$.  
A possible way this correspondence may work is as follows. 
In the (AP,AP) case, if we treat the black D4 solution (the blue region 
in the bottom half of Figure
\ref{fig-D4-AP-phase}), which is not related to YM4, as irrelevant, and focus on the solitonic D4 brane, its winding modes would appear to be light
around the temperature (\ref{winding-SD4}).
A Hagedorn transition parallel to the (P,AP) case might occur above this temperature and might continue to the confinement/deconfinement transition in YM4.
However, in order to investigate it further through gravity, we need to understand the gauge/gravity correspondence in the 0B frame as mentioned in footnote \ref{ftnt-0B}. 

\paragraph{Real time}
It would be important to explore what geometry corresponds to the
deconfinement phase in the real time formalism.
 An understanding of this would allow us to address
dynamical properties of the deconfinement phase, e.g. transport
properties. Since previous results in this area were based on the
black brane solutions, it would be important to see how well-known
results such as the viscosity bound \cite{Policastro:2001yc} can be
derived in our proposal\footnote{We would like to thank E. Kiritsis
  for emphasizing this issue to us.}.

\paragraph{AdS/CMT and chemical potential dependence}

The SS transition for D3 branes has been studied in the context of the
AdS/CMT correspondence to investigate the superconductor/insulator
transition in 2+1 dimension \cite{Nishioka:2009zj, Horowitz:2010jq}.
In these studies, a chemical potential for a $U(1)$ charge was
introduced and the phase structure involving this chemical potential
has been derived. It would be interesting to ask whether our proposal
has any bearing on these studies, e.g., whether the GL transition
analogous to the one discussed here can be a candidate for the
superconductor/insulator or some other transition.  A possible line of
investigation could be to study the chemical potential dependence of
$T_c$ for the GL transition {\em vis-a-vis} that for the SS transition
and see whether any qualitative differences appear\footnote{The
  chemical potential dependence for the confinement/deconfinement
  transition directly from gauge theory has been studied in a one
  dimensional model \cite{Morita:2010vi}.}.

\subsection*{Acknowledgement}
We would like to thank Avinash Dhar, Saumen Datta, Koji Hashimoto, Yoshimasa Hidaka,
Elias Kiritsis, Matthew Lippert, Shiraz Minwalla, Vasilis Niarchos,
Tadakatsu Sakai, Shigeki Sugimoto and Tadashi Takayanagi. We would also like to thank Ofer Aharony, Matthew Lippert, Shiraz
  Minwalla, Rob Myers and Vasilis Niarchos for a critical reading of the manuscript and for offering suggestions for improvement. 
 We would like to thank the organizers of the sixth regional String theory meeting
in Milos, Greece (June 2011) for hospitality during the
final stages of this work and for an opportunity to present this work.
  T.M. would like to
thank KEK for hospitality where part of the work was done. T.M. is
partially supported by Regional Potential program of the
E.U. FP7-REGPOT-2008-1: CreteHEPCosmo-228644 and by Marie Curie
contract PIRG06-GA-2009-256487.

\appendix

\section{Probe D7 brane solution in $T \gg T_{GL}$}
\label{app-probe}

In this appendix, we show the calculation of the probe
D7/$\overline{\rm D7}$ solution in the localized solitonic D3 brane
background (\ref{metric-LSD3}) in the $T \gg T_{GL}$ limit.

As we argued in section \ref{sec-chiral-SD3}, the D7/$\overline{\rm D7}$ would be localized at $t'=\beta'/2$.
Then, if the temperature is sufficiently high, we can use the Newtonian approximation \cite{Harmark:2004ws, Harmark:2002tr, Harmark:2003dg, Harmark:2003yz} around the position of the D7/$\overline{\rm D7}$ and
 the background metric in this region would be approximately described by \begin{align}
ds^2 = \a' \Biggl[&F^{-1/2}
\left( \sum_{i=1}^{3} 
dx_i^2 +(1+2\Phi)dx_4^2 \right) \nn
&+ F^{1/2} \left(1-\frac{1}{2}\Phi \right) \left(  du^2 +dt'^2+ u^2 d\Omega_{4}^2  \right)  \Biggr]  
, \nn
F= \frac{ d_3 \lambda_5}{\beta} &  \sum_n \left(  \frac{1}{u^{2}+(t'-n \beta')^{2}} \right)^2, \quad
\Phi=-\frac{\tilde{u_0}^4}{2} \sum_n \left(  
\frac{1}{u^2+(t'-n \beta')^2} \right)^2, \nn
 \quad e^{\phi}=\frac{\lambda_5}{2\pi N \beta} & .
\label{metric-bpsD3}
\end{align}
Here the sums in $F$ and $\Phi$ are  contributions of the mirror images of the localized SD3 on the $t'$-cycle and we have assumed $\Phi \ll 1$.
We consider the classical solution of the the D7/$\overline{\rm D7}$ brane on this background.
Since these branes would be localized at $t'=\beta'/2$, the non-trivial configuration of the branes would be $x_4(u)$ only. 
Then the induced metric on the D7/$\overline{\rm D7}$ becomes 
\begin{align}
ds^2_{D7} = \a' \Biggl[&F_0^{-1/2}
\left( \sum_{i=1}^{3} 
dx_i^2 \right)+ F_0^{1/2} \left(1-\frac{1}{2}\Phi_0 \right) u^2 d\Omega_{4}^2 \nn
& +\left(F_0^{-1/2} \left(1+2\Phi_0 \right) \dot{x}_4(u )^2+F_0^{1/2} \right)du^2  \Biggr] , \nn
F_0=\left.F  \right|_{t'=\beta'/2}&=\frac{  \lambda_4 L_4 \beta^3 }{64\pi^4 }   \frac{\sinh x -x }{x^{3} \cosh^2(x/2)},
\quad \Phi_0=-\frac{\pi^4 \lambda_4}{2^4 L_4^3 \beta}F_0, 
 \quad x=\frac{\beta u}{2\pi}   .
\end{align}
Here we have used $d_3=2$ and $\beta=(2\pi)^2/\beta'$. 
Now the DBI action\footnote{
We can verify the localization of the D7/$\overline{\rm D7}$ brane at $t'=\beta'/2$ by evaluating the fluctuation of $t'$ in this action.
} of the D7/$\overline{\rm D7}$ is given as
\begin{align}
S_{DBI}=& N_f T_7 \int d^8x e^{-\phi} \sqrt{\det g_{D7}} 
 \propto \int du \left(1-\Phi_0 \right)u^4 \sqrt{\left(1+2\Phi_0 \right)\dot{x}_4(u )^2+F_0} .
\label{action-DBI}
\end{align}
and the equation of motion for ${x}_4(u )$ from this action is
\begin{align}
\frac{d  }{d u } \left(  \frac{\left(1+\Phi_0 \right) u^4 \dot{x}_4(u )}{\sqrt{\left(1+2\Phi_0 \right)\dot{x}_4(u )^2+F_0}}  
\right)=0  .
\end{align}
Recall that the boundary condition for ${x}_4(u )$ is ${x}_4=L/2$ ($u \to \infty$).
(From now we restrict $x_4(u) \ge 0$ by assuming that the negative solution is obtained by $-x_4(u)$.)

From this equation, we can find two classical solutions.  One is a
constant solution: $x_4(u )=L/2$.  This solution corresponds to the
$\chi S$ line in Figures \ref{fig-probeD7} and \ref{fig-fene} (a).
The other solution is described by
\begin{align} 
x_4(u )=\int_{u_1}^{u} du' \left(1-\Phi_0(u') \right) \sqrt{ \frac{F_0(u')}{(u'/u_1)^8-1} } .
\label{x_4}
\end{align} 
Here we have assumed the existence of a turning point $u_1$, which satisfies
$x_4(u_1)=0$ and $\dot{x}_4(u_1)=\infty$.  This solution corresponds
to the $\chi SB$ profile in Figures \ref{fig-probeD7} and
\ref{fig-fene}. The value of $u_1$ can be determined by the condition
$x_4(u=\infty)=L/2$. We numerically\footnote{
In the numerical calculations, it is convenient to define the following dimensionless quantities: distance $L/L_4$, temperature $L_4 T/ \lambda_4$, cylindrical coordinate $L_4 u/ \lambda_4 $.
} find the relation between $u_1$
and $L$, which is plotted in Figure \ref{fig-fene} (b).  There is a
critical value $L_*(T)$, and, if $L$ exceeds it, no $\chi SB$ solution
exists whereas two solutions exist if $L< L_*$.  One solution (the
blue lines in Figure \ref{fig-probeD7} and \ref{fig-fene}) is always
unstable compared to another one (the red lines).

By comparing the DBI action (\ref{action-DBI}) for these two $\chi SB$
solutions and the $\chi S$ solution, we find that the $\chi S$
solution is stable for larger $L$.  A first order transition occurs at
$L=L_c(T)$ and the $\chi SB$ solution becomes stable below it.  See
Figures \ref{fig-fene} (a) and \ref{fig-tc}.  Note that $L_c(T)$ and
$L_*(T)$ are decreasing functions of temperature.  Thus the $\chi S$
phase is favoured at higher temperature, which is consistent with QCD.

One important question is whether chiral symmetry is restored at
$T=T_{GL}$. 
In our numerical calculation, we find that, 
if $L< L_c(T_{GL})=0.12 L_4$, the confinement/ deconfinement transition and the chiral symmetry restoration transition are separated as depicted in Figure \ref{fig-tc}. 
However, we have used the high temperature approximation in this
calculation and the result would not be reliable near $T_{GL}$.  In
addition, the background metric (\ref{metric-LSD3}) was derived using
an approximation which also does not work near this temperature. In
order to improve the result near $T_{GL}$, we may need precise
numerical study as in \cite{Kudoh:2004hs}.

\end{document}